\documentclass[fleqn,usenatbib]{mnras}
\usepackage{newtxtext,newtxmath}

\usepackage[T1]{fontenc}
\usepackage{ae,aecompl}

\usepackage{graphicx}	% Including figure files
\usepackage{amsmath}	% Advanced maths commands
\usepackage{amssymb}	% Extra maths symbols
\usepackage{color}
\title[]{New evidence for Dicke's superradiance in the 6.7 GHz methanol spectral line in the interstellar medium}
\author[F. Rajabi et al.]{F. Rajabi,$^{1}$\thanks{E-mail: f3rajabi@uwaterloo.ca}
M. Houde,$^{2}$\thanks{E-mail: mhoude2@uwo.ca}
A. Bartkiewicz,$^{3}$
M. Olech,$^{3}$
\newauthor M. Szymczak,$^{3}$
and P. Wolak$^{3}$
\\
$^{1}$Institute for Quantum Computing and Department of Physics and Astronomy, The University of Waterloo, 200 University Ave. West, \\Waterloo, Ontario N2L 3G1, Canada\\
$^{2}$Department of Physics and Astronomy, The University of Western Ontario, 1151 Richmond Street, London, Ontario N6A 3K7, Canada\\
$^{3}$Centre for Astronomy, Faculty of Physics, Astronomy and Informatics, Nicolaus Copernicus University,\\
Grudziadzka 5, PL-87100 Torun, Poland
}

\date{}

\pubyear{2018}

\begin{document}
\label{firstpage}
\pagerange{\pageref{firstpage}--\pageref{lastpage}}
\maketitle
% * <fereshte.rajabi@gmail.com> 2018-09-17T17:26:16.116Z:
%
% ^.
% Abstract of the paper
\begin{abstract}
We present new evidence for superradiance in the methanol 6.7 GHz spectral line for three different star-forming regions: S255IR-NIRS3, G24.329+0.144, and Cepheus A. Our analysis shows that some of the flux-density flares exhibiting fast rise times and asymmetric light curves reported in these sources can naturally be explained within the context of superradiance. When a threshold for the inverted population column density is exceeded in a maser-hosting region, the radiation mode switches from one regulated by stimulated emission (maser) to superradiance. Superradiance, as a more efficient energy release mechanism, manifests itself through strong bursts of radiation emanating from spatially compact regions. Elevated inverted population densities and the initiation of superradiance can be due to a change in radiative pumping. Here, we show that an increase in the pump rate and the inverted population density of only a factor of a few results in a significant increase in radiation. While the changes in the pump rate can take place over a few hundred days, the rise in radiation flux density when superradiance is initiated is drastic and happens over a much shorter time-scale. 
\end{abstract}

% Select between one and six entries from the list of approved keywords.
% Don't make up new ones.

%%%%%%%%################# Introduction   #########################
%%%%%%%%############################################################
\section{Introduction}
Dicke's superradiance (SR; \citealt{Dicke1954}) has been the subject of intensive research in the physics community for decades \citep{Gross1982,Benedict1996}. However, it had remained unknown to the astronomy community until its recent applications to the interstellar medium (ISM) \citep{Rajabi2016A,Rajabi2016B,Rajabi2016Thesis,Rajabi2017}. SR, a coherent radiation mechanism, can provide explanation for strong radiation bursts taking place over a wide domain of time-scales ranging from sub-seconds (\citealt{Mathews2017,Houde2018b}; \citealt*{Houde2018a}) to several years \citep{Rajabi2016B}.

Two of the necessary conditions for SR, i.e., population inversion and velocity coherence among emitters, motivated the search for SR in masing regions where these conditions are known to be met. Accordingly, evidence for SR was so far found for the methanol 6.7 GHz and water 22 GHz spectral lines in star-forming regions \citep{Rajabi2017} and for the OH 1612 MHz line in the circumstellar envelope and environment of evolved stars \citep{Rajabi2016B}. In all these cases, observed variations in the shape of the light-curves (e.g., damped oscillatory behavior) in some velocity components or significant flux density increases over relatively short time-scales are naturally accounted for within the context of SR. 

The build-up of coherence in SR necessitates an additional condition compared to masers. That is, the characteristic time-scale of SR $T_{\mathrm{R}}$ should be shorter than those of all non-coherent processes (e.g., the time-scale of collisions). For elongated cylindrical molecular ensembles, which we use to mathematically model SR samples that may get established in the ISM along the line-of-sight, $T_{\mathrm{R}}$ depends on the column density of the inverted population for a given spectral line. This thus implies the existence of a critical column density above which SR can ensue. It therefore follows that a transition from maser to SR radiation takes place when a critical inverted column density of partaking molecules is exceeded in the corresponding medium, as will be shown in Section \ref{sec:Results} \citep{Rajabi2016B,Rajabi2017}. Amplification by stimulated emission, the underlying mechanism for masers, does not require this condition as it does not involve entanglement between the atoms or molecules. Therefore, when the inverted population column density is less than the critical value (i.e., $nL<(nL)_{\mathrm{crit}}$, where $n$ and $L$ are the inverted population density and length of the column, respectively) masers can be a natural outcome. In masers, the intensity scales linearly with the number of emitters $N$ ($I_{\mathrm{m}} \propto N$, where $I_{\mathrm{m}}$ stands for the maser intensity). In contrast, with SR emitters act as a unit through the establishment of entangled states and the radiation rates are significantly enhanced. As a result, the radiation intensity at maximum scales with the square of the number of emitters ($I_{\mathrm{SR}} \propto N^2$) \citep{Dicke1954,Dicke1964}. SR therefore proves a more efficient mechanism to release the energy stored in an inverted medium through strong bursts of radiation over a variety of time-scales.

Variation in maser sources in the form of flares are known for OH, water, methanol, ammonia, and formaldehyde (\citealt{Gray2012} and references therein). The methanol maser sources are among the most intensively surveyed ones because of their high intensity that facilitates their study using smaller radio telescopes available for long-term monitoring programs. The $5_1 - 6_0\mathrm{A}^+$ transition at 6.668519 GHz is the strongest methanol maser line \citep{Menten1991}, and maser flares in this line were first reported by \citet*{Goedhart2004} toward a number of star-forming regions. The 6.7~GHz masers are believed to be pumped radiatively by infrared emission from warm dust \citep{Sobolev1997,Cragg2005} and, naturally, variations in the pump can cause variations in the maser flux density. Periodic, quasi-periodic, or aperiodic time-variability are reported for 6.7 GHz methanol flares (\citealt{Goedhart2004, Fujisawa2014a}; \citealt{Szymczak2015}; \citealt{Szymczak2018a}), with only a minority in the first two groups \citep*{Gray2018}. In some cases, the coexistence of periodic and aperiodic flares within the same gas volume of a few hundred au suggests local changes in the pump conditions \citep{Szymczak2015}. These flares seem to occur over a wide domain of time-scales, ranging from 23.9 days \citep{Sugiyama2017} to several hundred days \citep{Goedhart2004,Szymczak2018a}. Their profiles exhibit both symmetric and asymmetric features. For asymmetric cases, relatively short flux rise times are followed by slow decays. 

Different theories have been proposed to explain flaring in masers sources. These theories relate the observed time-variations to: changes in the background free-free emission \citep*{Van2009, Van2011}, variations in the infrared radiation field by accretion onto a young binary
system \citep{Araya2010} or stellar pulsation \citep{Inayoshi2013} or rotation of hot and dense material of the spiral shock wave (for periodic flares; \citealt{Parfenov2014}), variations due to heating dust and gas by magnetic reconnection from a central star \citep{Fujisawa2012, Fujisawa2014b}, time-variability resulting from rotation of the maser cloud (for quasi-periodic flares; \citealt{Gray2018}), superimposition of two or more clouds along the line-of-sight \citep{Deguchi1989, Boboltz1998}, and more. One common characteristics of all these models is that they explain flares using the maser mechanism, i.e., amplification through the stimulated emission process in the emitting gas of photons either spontaneously emitted at the maser frequency within the source or from background radiation. Here, we argue that the SR model provides a viable alternative for these cases.

In this paper, we apply the SR model to flaring events toward three different star-forming regions. In Section \ref{sec:Methods}, we give a short overview of our model and the observational data used for our analysis. We then present our results in Section \ref{sec:Results} and discuss how SR can reproduce the observed profiles and time-scales of flares for different velocity components in the aforementioned sources. Finally, we end with a summary and conclusion in Section \ref{sec:conclusion}.

%%%%%%%%###################### Methods ############################
%%%%%%%%############################################################
\section{Methods and data}\label{sec:Methods}
\subsection{Numerical methods}
The computation methods in this paper are similar to what was used in \citet{Houde2018b}, with the exception that in the present case the initiation of SR is accomplished through a pumping pulse at a shorter wavelength (as opposed to a triggering pulse at the frequency of the SR transition). Otherwise, the methodology below closely follows that found in \citet{Arecchi1970, Gross1982,Benedict1996,Houde2018b}.

The evolution of a group of inverted atoms/molecules interacting with a radiation field as a function of position $z$ and retarded time $\tau = t - z/c$ is given by the so-called Maxwell-Bloch equations:
\begin{eqnarray}
     \frac{\partial\hat{\mathbb{N}}}{\partial\tau} & = & \frac{i}{\hbar}\left(\hat{P}_0^+\hat{E}_0^+-\hat{E}_0^-\hat{P}_0^-\right)-\frac{\hat{\mathbb{N}}}{T_1}+\hat{\Lambda}_\mathbb{N} \label{eq:dN/dt} \\     
     \frac{\partial\hat{P}_0^+}{\partial\tau} & = & \frac{2id^2}{\hbar}\hat{E}_0^-\hat{\mathbb{N}}-\frac{\hat{P}_0^+}{T_2} \label{eq:dP/dt} \\
    \frac{\partial\hat{E}_0^+}{\partial z} & = & \frac{i\omega_0}{2\epsilon_0c}\hat{P}_0^-. \label{eq:dE/dt}
\end{eqnarray}
Here, $2\hat{\mathbb{N}}$ is the inverted population density, while $\hat{P}_0^+$ and $\hat{E}_0^+$ are, respectively, the amplitudes of the polarization in the medium and of the electric component of the radiation field. These equations describe a one-dimensional system that is suitable for the study of cylindrical samples\footnote{As will be discussed in Section \ref{subsec:S255}, upon the initiation of SR a region initially hosting a maser will break up into a large number of smaller entities where SR takes place. These smaller entities are what we refer to as ``samples.'' More precisely, the cylindrical structures considered here of length $L \gg \lambda$, with $\lambda$ the wavelength of radiation, are designated as ``large samples'' in the physics literature (as opposed to ``small samples'' where $L \ll \lambda$; for example, see \citealt{Gross1982}).} that would naturally form along the line-of-sight. In the derivation of equations (\ref{eq:dN/dt})-(\ref{eq:dE/dt}), we applied the slowly varying envelope and the rotating wave approximations \citep{Arecchi1970,Gross1982,Benedict1996}. Also, the resonant polarization and electric field vectors are given by 
\begin{eqnarray}
     \mathbf{\hat{P}}^\pm\left(z,\tau\right) & = & \hat{P}_0^\pm\left(z,\tau\right)e^{\pm i\omega_0\tau}\boldsymbol{\varepsilon}_d \label{eq:P_svea} \\
    \mathbf{\hat{E}}^\pm\left(z,\tau\right) & = & \hat{E}_0^\pm\left(z,\tau\right)e^{\mp i\omega_0\tau}\boldsymbol{\varepsilon}_d, \label{eq:E_svea}
\end{eqnarray}
where $\boldsymbol{\varepsilon}_d$ is the unit vector associated with the atomic/molecular transition at the angular frequency $\omega_0$ and of dipole moment $d$. 

As a result of non-coherent processes (e.g., collisional de-excitation), the inverted population density undergoes a decay over a time-scale $T_1$. Similarly, dephasing processes (e.g., elastic collisions or Doppler broadening) weaken the polarization over a time-scale $T_2$. On the other hand, some non-coherent processes, which are accounted by a pump term $\hat{\Lambda}_{\mathrm{N}}$, increase the inverted population density in the sample. Throughout this paper, we assume a pump function given by  
\begin{equation}
\hat{\Lambda}_{\mathrm{N}}\left(z,\tau\right) = \hat{\Lambda}_{\mathrm{0}} + \frac{\hat{\Lambda}_{\mathrm{1}}}{\cosh^{2}\left[\left(\tau-\tau_0\right)/T_\mathrm{p}\right]},\label{eq:pump}
\end{equation}
where $\hat{\Lambda}_{\mathrm{0}}$ is a constant pump rate, $\hat{\Lambda}_{\mathrm{1}}$ is the amplitude of a pump pulse of duration $T_{\mathrm{p}}$ and centered at the retarded time $\tau_0$.

The inverted population density is initially assumed zero across the sample and allowed to reach a constant value $\hat{\mathbb{N}}_0 = \hat{\Lambda}_0 T_1$ at $\tau=0$ in our analysis. At that time the polarization in the sample is given by  $\hat{P}_0^+ = Nd \sin\left(2/\sqrt{N}\right)/(2V)$, with $N$ the number of inverted molecules in the SR sample of volume $V$. We also assume $\hat{E}_0^+\left(z,\tau=0\right) = 0$.

Equations (\ref{eq:dN/dt})-(\ref{eq:dE/dt}) are solved numerically for the given initial conditions using a fourth-order Runge-Kutta method adapted from \citet{Mathews2017} to obtain $\hat{\mathbb{N}}\left(z,\tau\right)$, $\hat{P}_0^+(z,\tau)$, and $\hat{E}_0^+(z,\tau)$ for $0\le z\le L$ and $0\le \tau \le \tau_{\mathrm{max}}$ \citep{Houde2018b}. As previously mentioned, the SR samples are assumed to have a cylindrical shape of length $L\gg \lambda$, with $\lambda$ the wavelength of the radiation. For our analysis to be valid we require that the samples admit a Fresnel number of unity \citep{Gross1982,Rajabi2017}, i.e., their cross-section $A = \lambda L$. The outgoing radiation intensity is computed using $I_\mathrm{SR}\left(\tau\right) = c\epsilon_0|\hat{E}_0^+\left(z=L,\tau\right)|^2/2$. 

The SR fits in the next section are produced by adjusting the initial inverted population level $2\hat{\mathbb{N}}_0$, the SR sample length $L$, the time-scales $T_1$ and $T_2$, as well as the constant pumping term $\hat{\Lambda}_0$, and the amplitude $\hat{\Lambda}_1$ and duration $T_\mathrm{p}$ of the pumping pulse. 

%%%%%%%%%%%%%%%%%%%%%%%%%%%%%%%%%%%%%%
\subsection{Modelling of the SR phenomenon in maser regions}

In the analyses to follow, we will be dealing with maser-hosting environments, where an inversion of a molecular population is initially established with sufficient velocity coherence to support the existence of a corresponding maser radiation. One of our goals is to detail how such an environment can transition between modes where maser or SR radiation emanate from the enclosed molecular gas. To better understand the underlying processes it will be advantageous to first describe the parameters and time-scales characterizing SR by considering the simplest possible set of conditions. That is, we will consider a medium where the molecular population is inverted through a narrow pumping pulse propagating longitudinally along the symmetry axis of the resulting cylindrical SR sample. We also further assume that the relaxation and dephasing time-scales found in equations (\ref{eq:dN/dt}) and (\ref{eq:dP/dt}) are such that $T_1=T_2\equiv T^\prime$. We again emphasize that these idealizations will be relaxed when dealing with and analyzing the data in the sections to follow.    

Under these conditions, coherent interactions will dominate and SR ensue in a medium when the characteristic time-scale of SR 
\begin{equation}
T_{\mathrm{R}} = \tau_{\mathrm{sp}}\frac{8\pi}{3\lambda^2\left(nL\right)_{\mathrm{SR}}}
\label{eq:TR}
\end{equation}
is shorter than the time-scale of the non-coherent relaxation/dephasing processes $T^\prime$. In equation (\ref{eq:TR}) $\tau_{\mathrm{sp}}$ is the time-scale of spontaneous emission (i.e., the inverse of the Einstein coefficient of spontaneous emission) and $\left(nL\right)_{\mathrm{SR}}$ is the inverted population column density as SR is initiated. More precisely, SR will occur when $T_{\mathrm{R}} \ll T^\prime$ with the release of the energy stored in the inverted medium through a strong burst of radiation after a time delay  
\begin{equation}
\tau_{\mathrm{D}} \approx \frac{T_{\mathrm{R}}}{4}\left|\ln\left(\pi\sqrt{N_{\mathrm{SR}}}\right)\right|^2, \label{eq:TauD}
\end{equation}
where $N_{\mathrm{SR}}$ is the number of entangled molecules contained in the volume occupied by the SR sample. This delay, which also scales with $T_{\mathrm{R}}$, corresponds to the time required for establishing coherence within the medium. We therefore find a stricter condition $\tau_{\mathrm{D}} < T^\prime$, since for large samples with $N_{\mathrm{SR}} \gg 1$ it is found that $\tau_{\mathrm{D}} > T_{\mathrm{R}}$ \citep{Rajabi2016A}.

The combination of equations (\ref{eq:TR})-(\ref{eq:TauD}) and $\tau_{\mathrm{D}} < T^\prime$ reveals the existence of a critical inverted column density 
\begin{equation}
 \left(nL\right)_\mathrm{crit}\approx \frac{2\pi}{3\lambda^2}\frac{\tau_\mathrm{sp}}{T^\prime}\left|\ln\left(\pi\sqrt{N_{\mathrm{SR}}}\right)\right|^2 \label{eq:nL_crit}
\end{equation}
defining the threshold above which SR can proceed. More precisely, SR will be initiated whenever $\left(nL\right)_{\mathrm{SR}}>\left(nL\right)_\mathrm{crit}$ \citep{Rajabi2016B,Rajabi2017}. When this condition is met the SR system responds to the pumping pulse with the emission of a powerful burst of radiation characterized by a fast rise time and a series of oscillations (i.e., ringing) in its light curve, which are increasingly damped with lower values of $T^\prime$ (see Figure 2 in \citealt{Rajabi2016B}).  

The existence of a critical inverted column density (i.e., equation~\ref{eq:nL_crit}) makes it clear how a molecular population can proceed from a maser to a SR mode of radiation. More precisely, in an initially inverted population of column density below $\left(nL\right)_\mathrm{crit}$ the time-scale characterizing coherent interactions will be too long relative to $T^{\prime}$ and SR will not develop. It follows that the stimulated emission process will dominate along paths of sufficient velocity coherence throughout the region, resulting in maser radiation. On the other hand, if the column density of the inverted population is somehow increased and made to exceed $\left(nL\right)_\mathrm{crit}$, coherence is established within a time $\tau_\mathrm{D}$ and SR will overtake the emission process. A rise in the inverted population column density, which can stem from an increase in the corresponding density $n$ and/or the length $L$ of the paths exhibiting velocity coherence, are likely to be due to elevations in the pump.

%The Class II 
6.7 GHz methanol masers are believed to be primarily pumped by $20\,\micron-30\,\micron$ radiation from dust at temperatures $100\,\mathrm{K}<T_{\mathrm{d}}<200\,\mathrm{K}$ \citep{Sobolev1997}. A statistical study by \citet{Urquhart2015} confirmed that 99$\%$ of 6.7 GHz methanol sources in the Methanol Multibeam Survey \citep{Green2017} are associated with a dust clump. The changes in the characteristics of these clumps, such as elevations in temperature or density, can affect the efficiency of the pump as well as the pump photon flux. It follows that a more efficient or stronger pump could increase the inverted population column density in an initially maser-hosting region beyond the critical value and initiate the onset of SR. For one of the examples analyzed in this paper, flaring events in the 6.7 GHz methanol line are known to coincide with an IR outburst in the central source \citep{Caratti2017,Szymczak2018b}. We will model such outbursts with a pumping pulse of a given amplitude and duration and show how they can initiate SR. The shape of these pulses are assumed to be symmetric for simplicity (see eq. (\ref{eq:pump})), although asymmetric pump pulses could also be used. 

An important feature of the flaring events discussed in this paper is their significant flux density increases over short time-scales. For instance, in the case of the S255IR-NIRS3 flaring event, an increase in flux of more than a factor of 1000 with a rise time of less than 100 days was seen for some velocity components \citep{Szymczak2018b}. To explain such steep flux density variation within the context of maser theory, a significant increase in the pump rate over a short time-scale would be necessary. As will be shown in Section \ref{subsec:S255}, the observed sharp flux density rises find a natural explanation within the context of SR, and require an increase of only a factor of a few in the inverted population density over a time-scale that is significantly longer than that characterizing the sharp flux density increase. This behavior is accounted for by the fact that as the inverted column density crosses the threshold set by the critical value (equation~\ref{eq:nL_crit}) one sees the time-scale of radiation transit from that characterizing stimulated emission to that of SR, which depends inversely on $\left(nL\right)_\mathrm{SR}$ and is much shorter (see equation~\ref{eq:TR}). Furthermore, the intensity of radiation goes from maser mode $I_{\mathrm{m}} \propto N$ to a coherent one $I_{\mathrm{SR}} \propto N^2$. The combination of the reduced time-scale and increased radiation intensity are at the root of the observed sharp rises in flux.

%%%%%%%%%%%%%%%%%%%%%%%%%%%%%
\subsection{Observational data}

We have used the 6.7\,GHz methanol maser observations from the Torun monitoring program which started in 2009 \citep{Szymczak2018a, Szymczak2018b} and is being continued. The spectra of channel spacing of $0.044\mathrm{km~s}^{-1}$ and typical $3\sigma$ sensitivity of 0.8--1.2\,Jy  were obtained with the 32\,m dish in a frequency switching mode. The flux density calibration was better than 10~per~cent. Data were taken at least once a month but typically at irregular intervals of 5--10 days, sometimes with gaps of 3--4 weeks due to scheduling constraints. 

The sources S255IR-NIRS3, G24.329+0.144, and Cepheus A were selected for our analysis as some of their flux-density flares exhibit fast rises and asymmetric light curves, which are expected and predicted by SR, and can thus find a natural explanation within this context.

%%%%%%%%%%%%%%%%%%%%%%%%%%%%%%%%%%%%%%%%%%%%%%%%%%%%%%%%%%%%%%%%%%%%
\section{Results and discussion}\label{sec:Results}

In the next three subsections we present and model data for S255IR-NIRS3, G24.329+0.144, and Cepheus A with the SR model detailed in Section \ref{sec:Methods} for the general case when $T_1\neq T_2$. We only select a few velocity channels from spectral features that best show evidence for SR (not all spectral features and flares are consistent with SR), as our main goal is to establish its presence in the methanol 6.7 GHz line in star-forming regions. Detailed analyses yielding more comprehensive models will follow in future publications. 

%%%%%%%%%%%%%%%%%%%%%%%%%%%%%%%%%%%%%%%%%%%%%%%%%%%%%%%%%%%%%%%%%%%%
\subsection{S255IR-NIRS3}\label{subsec:S255} \noindent
The S255IR-NIRS3 star-forming region (also known as S255IR-SMA1
or G192.6000.048) is a massive young stellar object located $1.78$ kpc away \citep{Burns2016}. \citet{Menten1991} reported the first detection of a 6.7 GHz methanol maser in this source. Follow-up monitoring observations \citep{Goedhart2004, Sugiyama2008a} showed only moderate flux-density variations (i.e., less than $25-30\%$) until 2015, when \citet{Fujisawa2015} detected a new strong methanol flare. \citet{Szymczak2018b} conducted an 8.5-year monitoring program of this source between 2009 and 2018 and compiled their data with previous observations to produce a 27-year long light curve of the 6.7 GHz line in S255IR-NIRS3 (see Fig. 2 in \citealt{Szymczak2018b}). These observations revealed several emission features in a velocity range spanning $0.6~\mathrm{km~s}^{-1}$ to $8.7~\mathrm{km~s}^{-1}$. 

\begin{center}
   \begin{figure}
        \includegraphics[width=\columnwidth]{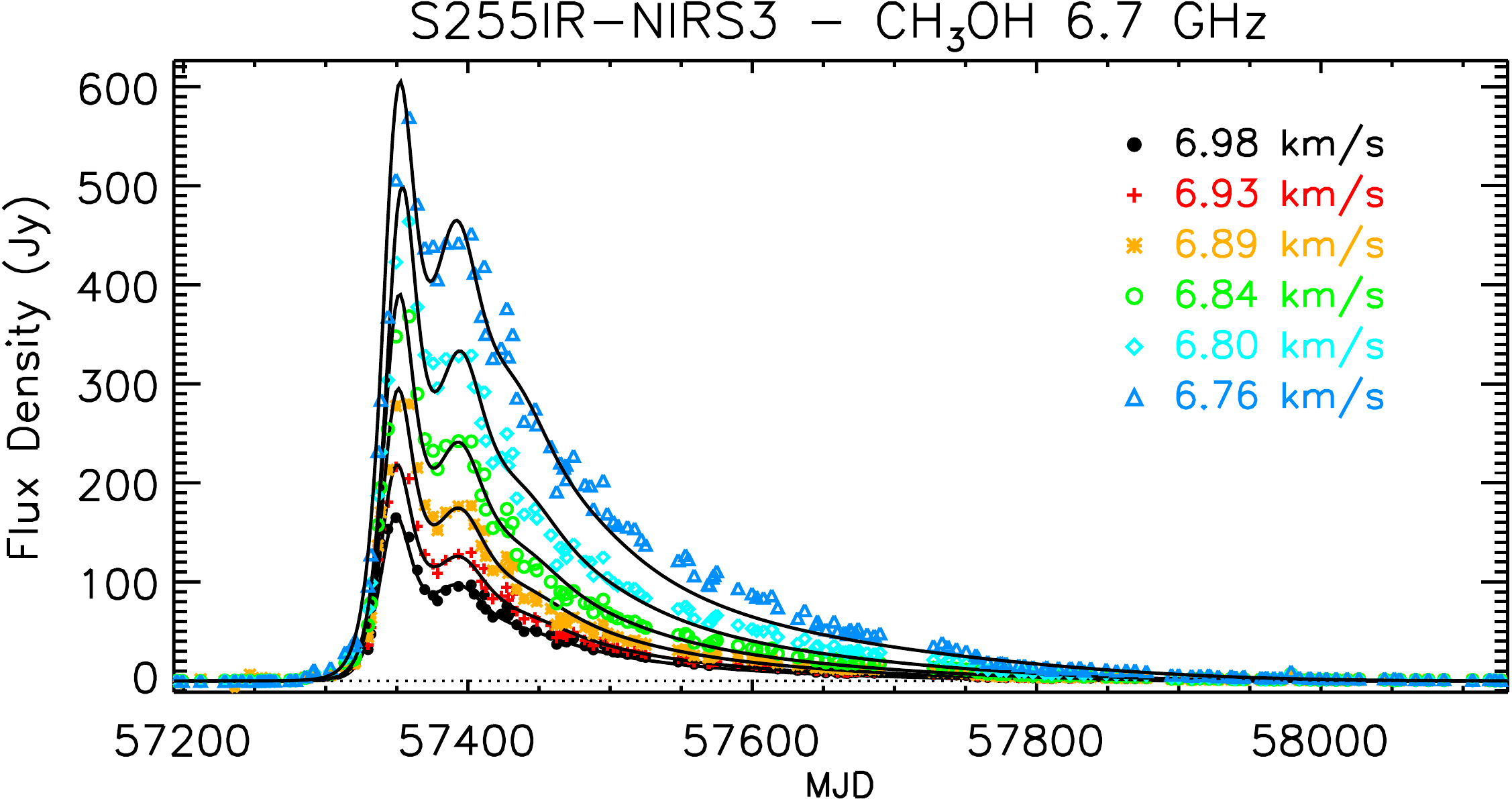}
        \caption{SR models (solid curves) for the S255IR-NIRS3 flaring event occurring between MJD 57200 and MJD 58126 for the $v_{\mathrm{lsr}}=6.76$ to $6.98~\mathrm{km~s}^{-1}$ velocity channels (symbols, using the legend on the top right). The SR fits are produced using SR samples of length $140~\mathrm{au} \leq L \leq 162$~au from the lowest to the strongest flux densities containing $\sim 10^{19}-10^{20}$ inverted and entangled molecules. The relaxation and dephasing time-scales are $T_1=1.64 \times 10^{7}$~s and $T_2 =1.55 \times 10^6$~s, respectively, for all velocities, while the column density of the inverted population $(nL)_\mathrm{SR}\simeq6 \times 10^{3}~\mathrm{cm}^{-2}$ sets the time-scale of the burst. The SR model flux densities are scaled to the data.}
        \label{fig:S255Fit}
    \end{figure}
 \end{center}

In Figure \ref{fig:S255Fit}, we show light curves of six velocity channels of $0.044~\mathrm{km~s}^{-1}$ in width, i.e., from $6.76~\mathrm{km~s}^{-1}$ to $6.98~\mathrm{km~s}^{-1}$, between MJD 57200 and MJD 58126. The symbols are for the data (using the legend on the top right of the graph) while the solid curves are for our corresponding SR model fits as described in the following. These velocity slices reside on the high-velocity edge of a spectral feature centered at approximately $6.4~\mathrm{km~s}^{-1}$ that appeared around MJD 57000 and reached its peak flux density some $\sim300$ days later. For example, at $6.76~\mathrm{km~s}^{-1}$ the flux density peaked at approximately 570 Jy while the pre-burst level was well below 1 Jy, yielding an increase in flux of the order of 1000 in less than 100 days. After this phase the flux density showed a rapid decay followed by a second maximum approximately 50 days later. The radiation level then decayed to pre-burst values over a time span of 720--740 days. 

Interestingly, the onset of flux-density bursts in this source coincides with episodic accretion on the central source, as inferred through near-infrared imaging \citep{Caratti2017}. These observations revealed an increase in IR brightness of 3.5 mag in the H (1.65 $\mu$m) and 2.5 mag in the K (2.16 $\mu$m) bands in November 2015. It was further determined from light echo that this IR burst started on MJD 57188 (i.e., mid-June 2015; \citealt{Caratti2017,Szymczak2018b}). The accretion was also later confirmed through detection of a burst in radio continuum at 6--46 GHz that started after MJD 57579 \citep{Cesaroni2018}.  This increase in IR luminosity is expected to elevate the population of inverted methanol molecules and consequently the flux density of the 6.7 GHz line. The asymmetry in the light curves, as seen in Fig. \ref{fig:S255Fit}, is another aspect of this flaring event that needs to be addressed.

\begin{center}
   \begin{figure}
        \includegraphics[width=\columnwidth]{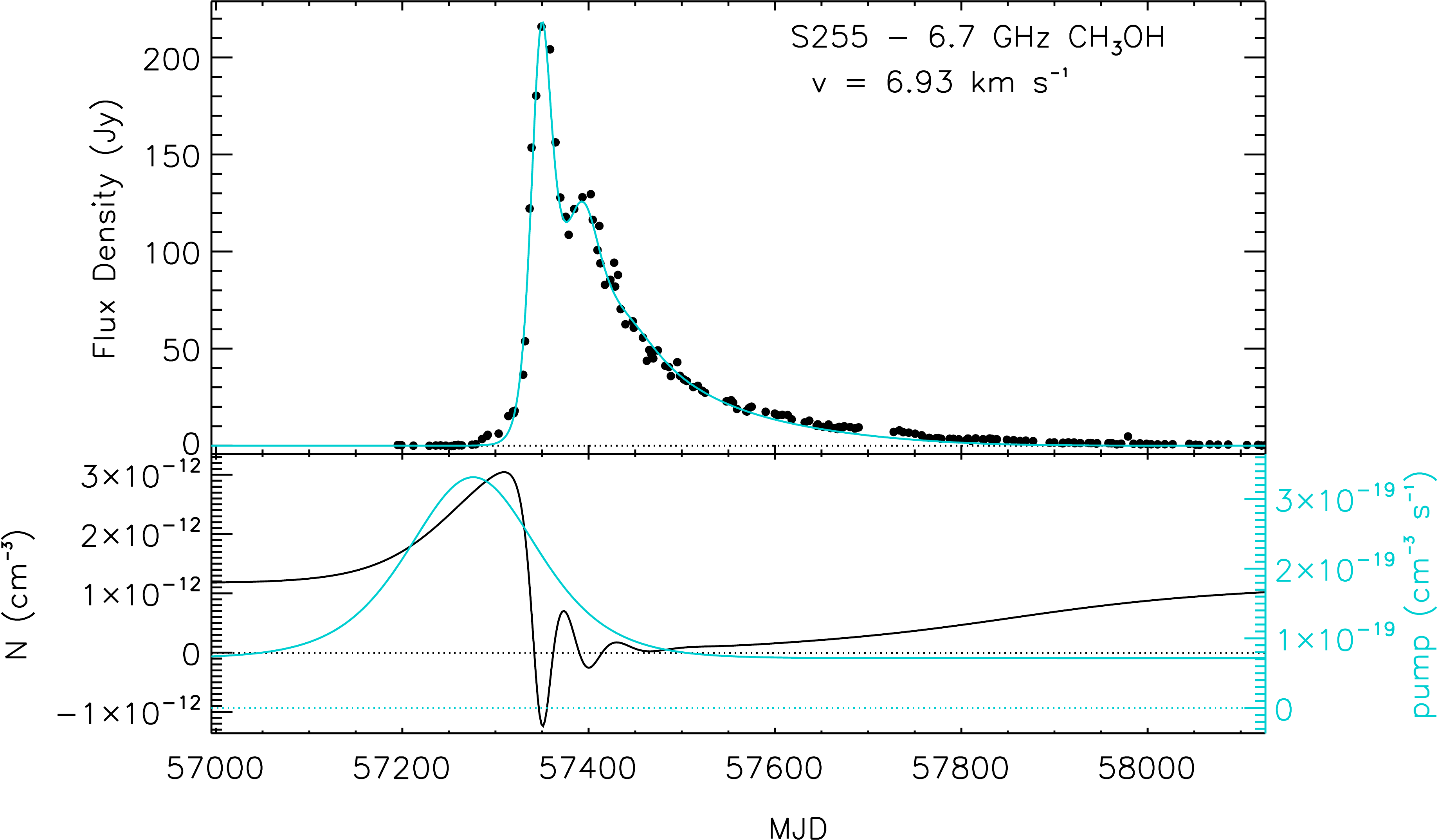}
        \caption{SR model for S255IR-NIRS3 flare at $v_{\mathrm{lsr}} = 6.93~\mathrm{km~s}^{-1}$. Top: The black dots are for the data and the solid cyan curve for the SR model fit as a function of retarded time. Bottom: The solid black and cyan curves, respectively, show the temporal evolution of the inverted population density and the pumping rate. The SR fit is produced using a single SR sample of length $L = 140$~au composed of $N_\mathrm{SR} \simeq 6\times 10^{19}$ inverted and entangled methanol molecules. The relaxation and dephasing time-scales are $T_1=1.64 \times 10^{7}$~s and $T_2 =1.55 \times 10^6$~s, respectively, while the inversion level prior to the appearance of the pump pulse corresponds to approximately $0.1~\mathrm{cm}^{-3}$ for a molecular population spanning a velocity range of $1~\mathrm{km~s}^{-1}$. The column density of the inverted population is $(nL)_\mathrm{SR}=6.4 \times 10^{3}~\mathrm{cm}^{-2}$ and the SR flux density is scaled to the data.}
        \label{fig:S255FitPump}
    \end{figure}
 \end{center}

In Figure \ref{fig:S255FitPump} we focus on the $v_{\mathrm{lsr}} = 6.93~\mathrm{km~s}^{-1}$ velocity channel to detail our models and the information they provide. We first account for the existence of maser radiation by setting an initial level for the inverted population density. For this, we assume an inversion density $n = 0.1~\mathrm{cm}^{-3}$ for a molecular population spanning a velocity range of $1~\mathrm{km~s}^{-1}$, or $\sim 10^4$~Hz at 6.7~GHz. Given the duration of the burst of $\sim10^7$~s the corresponding spectral extent will be of the order of $10^{-7}$~Hz, which translates to an inverted population density of $\sim 10^{-12}$ cm$^{-3}$ for a single SR sample. We therefore used this value as a starting point and refined it to $\hat{\mathbb{N}}_0 = 1.17 \times 10^{-12}$ cm$^{-3}$ during the fitting exercise. As was discussed in Section \ref{sec:Methods}, this initial level of inversion is given by $\hat{\mathbb{N}}_0 = \hat{\Lambda}_0 T_1$, with $\hat{\Lambda}_0$ and $T_1$ the constant pump rate and relaxation time-scale, respectively. 

The SR burst is initiated by the arrival of a symmetric pump pulse, given by equation (\ref{eq:pump}), of amplitude $\hat{\Lambda}_1 = 2.6\times 10^{-19}$ cm$^{-3}$s$^{-1}$ and duration $T_{\mathrm{p}} = 8.1\times 10^6$~s. In the bottom panel of Figure \ref{fig:S255FitPump} we show the total pump rate $\hat{\Lambda}_\mathrm{N}$, which is composed of the constant level $\hat{\Lambda}_0$ and the pulse of amplitude $\hat{\Lambda}_1$, as a function of time (cyan curve; using the vertical scale on the right). Also shown in the bottom panel is the temporal evolution of the population inversion density (black curve) at the end-fire (i.e., $z = L$) of the methanol sample (vertical scale on the left), while in the top panel the corresponding model flux density (cyan curve), as fitted and scaled to the data (black dots), is given. It is seen that after the arrival of the pump pulse the inverted population density of the sample increases until it reaches approximately three times pre-pump values, at which point SR is initiated. The energy stored in the molecules partaking in the cooperative effect is released through a strong burst exhibiting a sharp rise in flux density at the end-fire. As expected, the flux and inverted population density are negatively correlated, and the peak flux density coincides with a dip in $\hat{\mathbb{N}}$, where inverted population density becomes negative. That is, there are then more molecules in the ground state than the excited level. Photons emitted from $z<L$ in the sample are then absorbed at the end-fire ($z=L$), bringing the inverted density to positive values with the subsequent release of energy through a secondary maximum in the flux density. This is an example of the so-called SR ringing effect. After this phase, the non-coherent/dephasing processes that work against SR obstruct further development of cooperative emission and the inversion level oscillates/decays and eventually increases again under the action of the constant pump term until it reaches the pre-burst steady state value. Accordingly, the flux density also decays gradually to its pre-burst value. 

It is important to note that, in Figure \ref{fig:S255FitPump}, while the pump pulse lasts for more than 200~days or so, the rise in flux density when SR is initiated is drastic and happens over a much shorter time-scale (i.e, an increase of more than 200~Jy in about 50~days). This significant increase in radiation rate takes place despite the fact that the pump rate and the inverted population density are only elevated by factors of a few. Furthermore, the symmetry of the pump pulse is not transmitted to the radiation burst, which displays a strong asymmetry resulting from the characteristic response of the SR sample to the excitation.   

The SR model is a good fit to the detected flare profile and yields an inverted column density $(nL)_\mathrm{SR}=6.4 \times 10^{3}~\mathrm{cm}^{-2}$, which corresponds to $L = 140$~au with our assumed inverted population density. The radius of the cylindrical SR sample is found to be $5.5\times10^7~\mathrm{cm}$ (i.e., $3.7 \times 10^{-6}$~au), while $\simeq6\times10^{19}$ molecules were entangled and participated in the SR emission process. The relaxation and dephasing time-scales were set to $T_1=1.64 \times 10^{7}$~s and $T_2 =1.55 \times 10^6$~s, respectively. Equating the latter to the mean time between elastic collisions implies a gas density of approximately $10^5~\mathrm{cm}^{-3}$. Assuming a methanol abundance of $\sim10^{-5}$ \citep{Cragg2005} would imply an inversion level of about 0.1. We note, however, that the SR model only specifies the inverted column density $(nL)$ and not the density nor the sample length independently, as the value for $(nL)$ is set by the time-scale characterizing the data. It follows that a reduction in the inverted density could be offset by the opposite change in $L$ in order to keep $(nL)$ (and the fit to the data) unchanged. In other words, our model can readily be adjusted to a range of inverted population densities and inversion efficiencies by accordingly changing the length of the SR sample. 
%\textbf{For instance one can think of SR samples of a few tens of au long similar to masers while a more efficient pump results in a higher $n$ and sets the radiation mode to SR. It should be also noted that the small cross-sectional area of a single SR sample on the order of $10^{-6}$~au at a distance of 1.78 kpc corresponds to an angular spot size that is well beyond resolution limits of current VLBI techniques.}     
 
Our calculations also reveal that our SR sample radiates with an integrated flux density of $3.7\times 10^{-34}~\mathrm{erg~s}^{-1}\mathrm{cm}^{-2}$ (or $3.7\times 10^{-4}~\mathrm{Jy}$ for a spectral width of $\sim10^{-7}$~Hz) at a distance of 1.8~kpc. In order to match the observed flux density of $\sim 200$ Jy, a group of $\sim 10^6$ SR samples should erupt simultaneously. Considering the compact size of a single SR sample ($\sim 10^{-6}$~au), a maser region with a spot size of, for example, $\sim 10$~au could easily host the needed number of SR samples. In other words, only a small fraction of the maser region's volume needs to see its inverted population column density $\left(nL\right)_\mathrm{SR}$ exceed the critical threshold $\left(nL\right)_\mathrm{crit}$ for the total radiation to match the observed flux density \citep{Rajabi2016B,Rajabi2017,Houde2018c}. We note that although the small cross sectional area of a single SR sample would be unresolved even with the longest VLBI baselines, it is the spot size of the inverted region hosting the many SR samples that is amenable to detection during observations. Therefore, as our scenario for explaining the flares is based on the transition of an inverted region from a maser regime to one supporting SR, we expect an SR source to be resolvable if the initial maser source was in the first place since they both originate from the same region. It also follows that the angular scale covered by the emerging SR radiation (i.e., that resulting from all of the superradiance samples) will be determined by the geometry of the inverted region \citep{Gross1982}, as is the case for the astronomical maser \citep{Houde2018b}.

%\textbf{It should be also noted that in spite of small cross sectional area of a single SR sample, which is well beyond resolution limit of VLBI, a group of $10^6$ SR samples erupting simultaneously, form an au size spot size that might be resolved through interferometry techniques.}

The SR models for the other velocity channels shown in Figure \ref{fig:S255Fit} are very similar to the one for $v_{\mathrm{lsr}} = 6.93~\mathrm{km~s}^{-1}$ we just described and found in Figure \ref{fig:S255FitPump}. More precisely, $T_1$ and $T_2$ are the same at all velocities while the pump levels vary by less than 1\%. The parameter that seems to account for the changes in the appearance of the light curves is the length of the SR samples, which goes from 140~au to 162~au as one moves from the lowest to the strongest intensities, as could be expected.
%%%%%%%%%%%%%%%%%%%%%%%%%%%%%%%%%%%%%%%%%%%%%%%%%%%%%%%%%%%%%%%%%%% 
%\subsection{The G24.329+0.144 star-forming region}\label{subsec:G24}
\subsection{G24.329+0.144}\label{subsec:G24} \noindent
G24.329+0.144 (loosely referred to as W42) is a high mass star-forming region near the H\,{\sevensize II} complex W42 located at a distance of 7.5 kpc \citep{Green2011}. There are two other nearby methanol sources, G24.494-0.039 and G24.790+0.083, which together with G24.329+0.144 form an expanding ring-like structure \citep{Caswell2011}. The morphology of this source was studied using the European VLBI Network (EVN) in June 2009 \citep*{Bartkiewicz2016} and a weak emission at 6.7 GHz line was reported, corresponding to a maser brightness temperature $<2\times 10^7$\,K. The comparison to detected flux levels from previous observations with the EVLA suggested strong variability in this source. \citet{Szymczak2018a} monitored G24.329+0.144 between MJD 55011 to MJD 56386, and detected a synchronized outburst in all spectral features. 

\begin{center}
   \begin{figure}
        \includegraphics[width=\columnwidth]{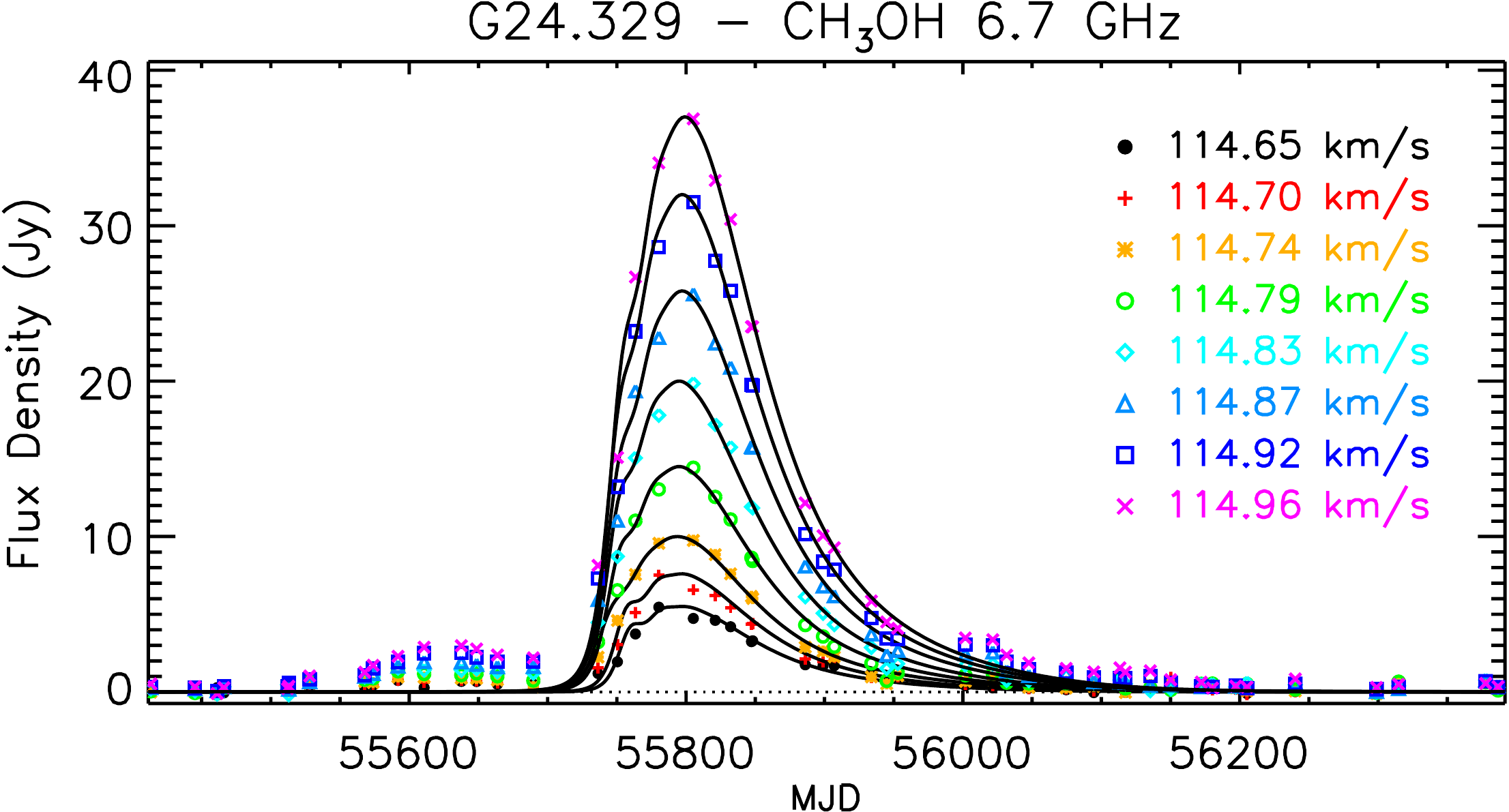}
        \caption{SR models (solid curves) for the G24.329+0.144 flaring event observed between MJD 55700 to MJD 56200 for eight velocity channels between $114.65~\mathrm{km~s}^{-1}\le v_{\mathrm{lsr}}\le 114.96~\mathrm{km~s}^{-1}$ (symbols, using the legend on the top right). The SR models are produced using SR samples of length $487~\mathrm{au} \leq L \leq 513$~au from the lowest to highest flux density. The relaxation and dephasing time-scales for all velocity channels are $T_1 = 1.7 \times 10^7$~s and $T_2 = 5.43 \times 10^5$~s, respectively. The column density of the inverted population is $1.16\times 10^4$~cm$^{-2}$. The SR model flux densities are scaled to the data.}
        \label{fig:G24p329Fit}
    \end{figure}
 \end{center}

In Figure \ref{fig:G24p329Fit}, we show the light curves of eight $0.044~\mathrm{km~s}^{-1}$ wide velocity channels at $114.65~\mathrm{km~s}^{-1}\le v_{\mathrm{lsr}}\le 114.96~\mathrm{km~s}^{-1}$ (symbols, using the legend on the top right). These velocity channels lie on the leading edge of a spectral feature centered at approximately $115.1~\mathrm{km~s}^{-1}$. Although two flares are present in the light curves, we focus on the second and strongest that took place between MJD 55700 and MJD 56200. The black solid curves superposed on the data correspond to our SR model fits. As can be seen in the figure, the flux density of different velocity channels rose over a time-scale of approximately 100~days to their peak values appearing around MJD 55805. This was then followed by a relatively slow decay in flux density over the next $\sim$200 days. The global variations in this source were thought to be associated with changes in the infrared pump photon rates while some features exhibited anti-correlation between their line width and intensity \citep{Szymczak2018a}, which is a characteristic of unsaturated amplification \citep{Goldreich1974}. In some velocity components up to a factor of $\sim$ 57 increase in flux density was reported (see \citet{Szymczak2018a} Section 4.3). To explain such a rise in flux density within the context of maser theory, the optical depth of the emitting gas should increase by a factor of 4 or more, assuming an unsaturated region, over the time-scale of a few months \citep{Szymczak2018a}.   

\begin{center}
   \begin{figure}
        \includegraphics[width=\columnwidth]{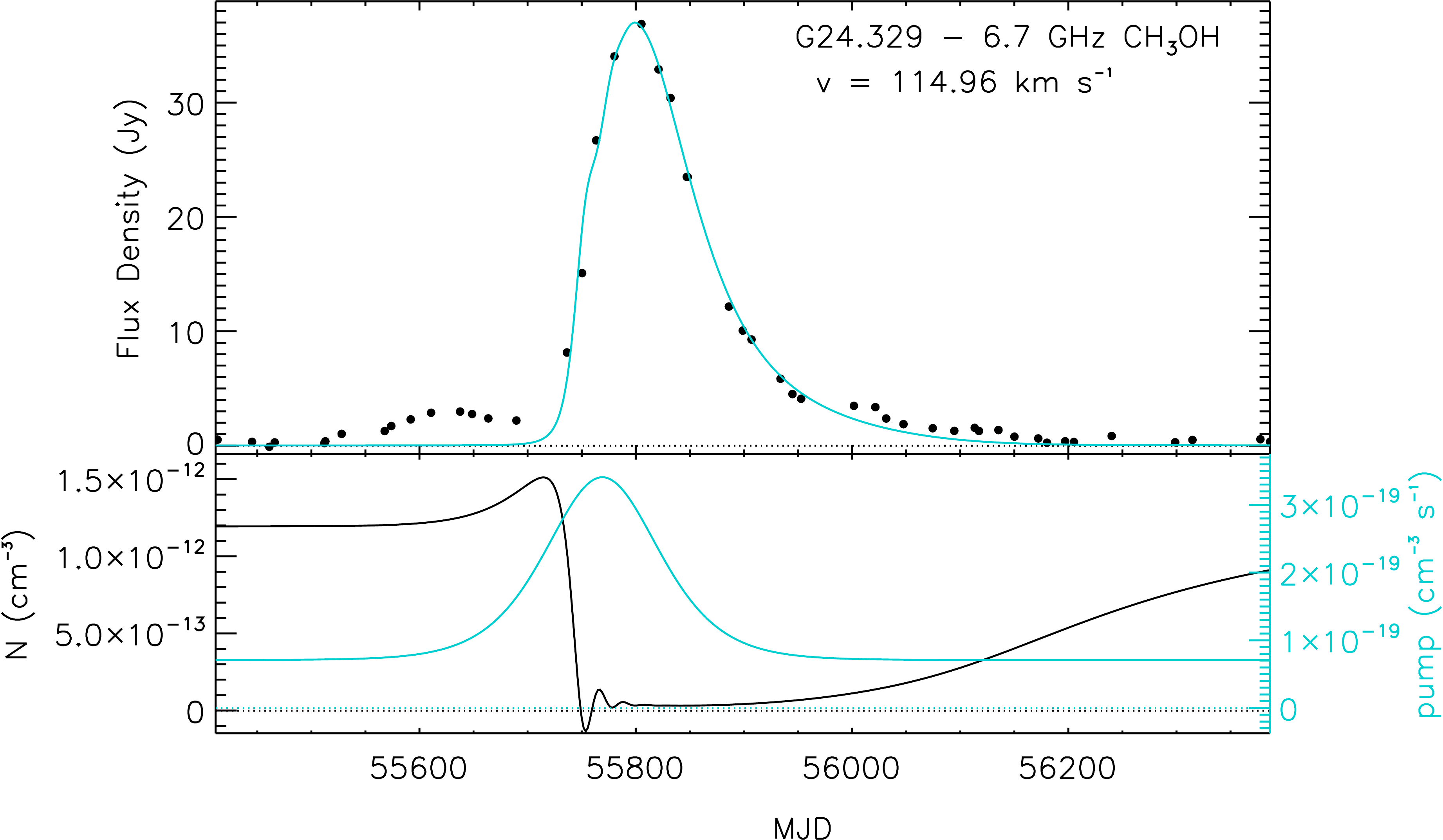}
        \caption{Same as Figure \ref{fig:S255FitPump} except for the G24.329+0.144 flare at $v_{\mathrm{lsr}} = 114.96~\mathrm{km~s}^{-1}$. The SR fit is produced using a SR sample of length $L = 513$~au containing $N_\mathrm{SR}\sim 4 \times 10^{20}$ excited methanol molecules, while the relaxation and dephasing time-scales are $T_1 = 1.7 \times 10^7$~s and $T_2 = 5.43 \times 10^5$~s, respectively. Prior to the arrival of the pump pulse, we assume an inversion level corresponding to approximately $0.1~\mathrm{cm}^{-3}$ for a molecular population spanning a velocity range of $1~\mathrm{km~s}^{-1}$. The SR model fit yields an inverted column density $\left(nL\right)_{\mathrm{SR}} = 1.16\times 10^4$~cm$^{-2}$. The flux density of the SR fit is scaled to the data.}
        \label{fig:G24p329FitPump}
    \end{figure}
 \end{center}

In Figure \ref{fig:G24p329FitPump}, as was the case for Figure \ref{fig:S255FitPump}, we select a velocity channel, here $v_{\mathrm{lsr}} = 114.96~\mathrm{km~s}^{-1}$, to discuss the SR model's fitting parameters. During the fitting process, we started with the same inverted population density (i.e., $n = 0.1~\mathrm{cm}^{-3}$ over a velocity width of $1~\mathrm{km~s}^{-1}$) as in the previous case. The flare spectral width of $\sim 10^{-7}$~Hz corresponds to an inversion level $\sim 10^{-12}~\mathrm{cm}^{-3}$, which during the fitting exercise was adjusted to $\hat{\mathbb{N}}_0 = 1.26 \times 10^{-12}$ cm$^{-3}$, a level slightly higher than what was used for the S255IR-NIRS3 model. The flare exhibits a relatively more symmetric light curves than the S255-NIRS3 burst, and was accounted for by a shorter dephasing time-scale $T_2 = 5.43 \times 10^5$~s. The relaxation time-scale was set to $T_1 = 1.77 \times 10^7$~s and, accordingly, a constant pump rate $\Lambda_0 = 7.1 \times 10^{-20}~\mathrm{cm}^{-3}~\mathrm{s}^{-1}$ was needed to produce the aforementioned initial population inversion level $\hat{\mathbb{N}}_0$. The pump pulse parameters were $\Lambda_1 = 2.7 \times 10^{-19}~\mathrm{cm}^{-3}~\mathrm{s}^{-1}$ and $T_{\mathrm{p}} = 6 \times 10^{6}$~s, once again a time-scale significantly longer than that over which the flux density rises at the beginning of the burst. The resulting SR fit gives a population inverted column density of $1.16\times 10^{4}$~cm$^{-2}$, almost twice higher than what we found for S255IR-NIRS3. This inverted column density corresponds to a SR sample length $L = 513$~au, and subsequently, a cross-sectional radius $w=1.05\times10^8$~cm (or $7 \times 10^{-6}$~au). For the given densities, a single SR sample is composed of $4 \times 10^{20}$ excited and entangled methanol molecules, which are responsible for a peak integrated flux density $1.13\times 10^{-33}~\mathrm{erg~s}^{-1}\mathrm{cm}^{-2}$ (or $1.13\times 10^{-3}$~Jy for a bandwidth of $10^{-7}$~Hz) at 7.5~kpc.    

The SR model fits for all velocity channels in Figure \ref{fig:G24p329Fit} employ the same inverted population density, $T_1$, $T_2$, and identical pump rate parameters. The only difference in the fitting parameters was for the SR sample lengths, where $488~\mathrm{au} \leq L \leq 514~\mathrm{au}$ as we moved from the lowest to highest flux density.

%%%%%%%%%%%%%%%%%%%%%%%%%%%%%%%%%%%%%%%%%%%%%%%%%%%%%%%%%%%%%%%%%%%%
%\subsection{The Cepheus A star-forming region}\label{subsec:cepA}
\subsection{Cepheus A}\label{subsec:cepA} \noindent
Cepheus A (also known as Cep A or G109.871+2.114) is a massive star-forming region located at a distance of $\sim0.7$ kpc \citep{Moscadelli2009}. \citet{Menten1991} first reported the detection of the 6.7 GHz line in this source with a flux density of 1420 Jy at $-6~\mathrm{km~s}^{-1} < v_{\mathrm{lsr}} < -1~\mathrm{km~s}^{-1}$. The strongest feature at $v_{\mathrm{lsr}} = -2.1~\mathrm{km~s}^{-1}$ subsequently dropped in flux density from 1420 Jy to 815 Jy over a timespan of 8.1 year \citep*{Szymczak2014}. Follow-up observations by \citet{Sugiyama2008b} identified five spectral features, where the redshifted components at -1.9 and $-2.7~\mathrm{km~s}^{-1}$ showed 50$\%$ decrease in flux over 81 days. The blueshifted features at -3.8, -4.2, and $-4.9~\mathrm{km~s}^{-1}$ were synchronized but anti-correlated with the redshifted ones, and exhibited an increase in flux within 30 days. This anti-correlation invalidated the association of methanol maser variabilities with changes in the collisional excitations by shock waves in this source \citep{Szymczak2014}. Observations with the JVN (Japanese VLBI Network) located the methanol masers in a filamentary arc-like structure of $\sim$1400 au near the thermal jet Cep A HW2 \citep{Sugiyama2008b}. The HW2 central object is a protostar with a mass of 18 M$_{\odot}$ \citep{Jimenez2009} surrounded by a circumstellar disk of 330 au in radius in dust and $580$ au in gas \citep{Patel2005}. This object is a strong radio continuum source \citep{Hughes1984} that is thought to be powering the region \citep{Rodriguez1994}. The molecular outflow associated with HW2 shows a complex morphology with high-velocity ($v>500~\mathrm{km~s}^{-1}$) ionized jet and a slow ($v = 10-70~\mathrm{km~s}^{-1}$) wide-angle outflow \citep{Torrelles2011}. 
\begin{center}
   \begin{figure}
        \includegraphics[width=\columnwidth]{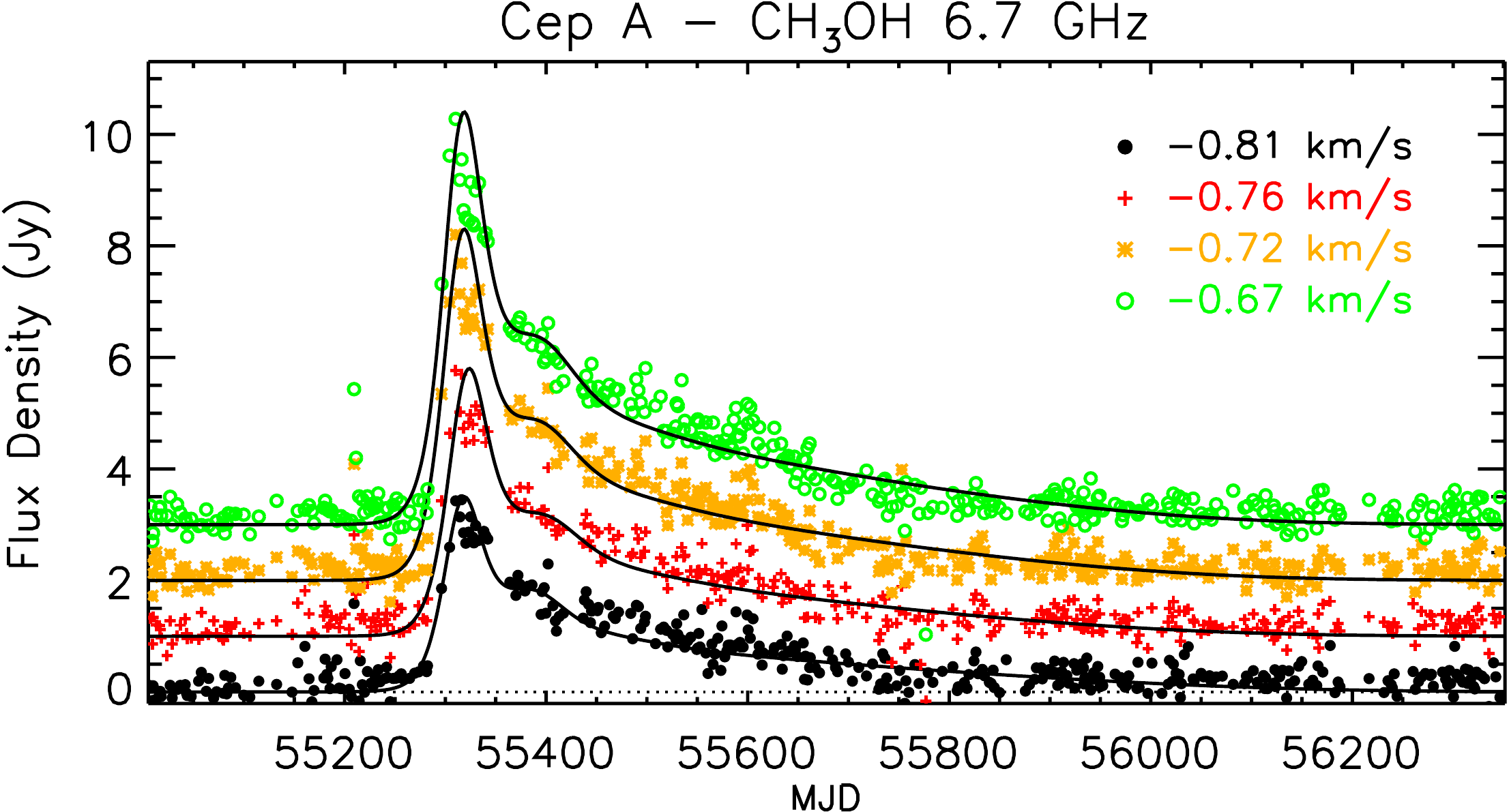}
        \caption{SR models (solid curves) for the Cep A flaring event detected between MJD 55007 to MJD 56346 for the $v_{\mathrm{lsr}} = -0.81, -0.76, -0.72, -0.67~\mathrm{km~s}^{-1}$ velocity channels (symbols, using the legend on the top right). The flux density levels are shifted by 0, +1, +2, and +3 Jy from the lowest to the highest velocity channel to better distinguish between the different light curves. The SR fits are produced using SR samples of length $L = 125$~au except at $v_{\mathrm{lsr}} = -0.81~\mathrm{km~s}^{-1}$ for which $L = 129$~au. The SR samples are composed of $\sim 10^{19}-10^{20}$ inverted and entangled molecules. The relaxation and dephasing time-scales are, respectively, $T_1 = 9.99\times 10^6$~s, $T_2 = 2.72\times 10^6$~s in all cases. The SR fits yield inverted column densities of $\simeq 4 \times 10^{3}$~cm$^{-2}$. The SR model flux densities are scaled to the data.}
        \label{fig:CepAFit}
    \end{figure}
\end{center}

\citet{Szymczak2014} monitored 6.7~GHz methanol masers in Cep A HW2 with the Torun 32 m telescope for a period of 1340 days. They also found synchronized and anticorrelated flux density variations between two blueshifted and one redshifted features over 30$\%$ of their observation period. Beginning on MJD 56056, periodic time-variability over time-scales of 84 to 87 days was observed in two spectral features. In Figure \ref{fig:CepAFit} we show a portion of the observations of \citet{Szymczak2014}, taken between MJD 55007 to MJD 56346, for four $0.044~\mathrm{km~s}^{-1}$ wide velocity channels located at $v_{\mathrm{lsr}} = -0.81, -0.76, -0.72, -0.67~\mathrm{km~s}^{-1}$ (symbols, using the legend on the top right). The flux density levels are shifted by 0, +1, +2, and +3 Jy from the lowest to the highest velocity channel to better distinguish between the different light curves. The solid black curves are for the corresponding SR model fits which will soon be discussed. The strongest increase in flux density occurs in the $v_{\mathrm{lsr}} = -0.76~\mathrm{km~s}^{-1}$ channel, where on MJD 55316 the flux density reached $\sim 29$ times the pre-burst level over a time-scale of approximately 90 days. The other three channels at $v_{\mathrm{lsr}} = -0.81, -0.72, -0.67~\mathrm{km~s}^{-1}$ also show an increase in flux by factors of $\sim$ 21, 16, and 19, respectively, on the same timespan. After this phase, the intensities decayed gradually to their pre-burst levels over a period of approximately three years. Although the peak flux density in Cep A is significantly lower than those of the previous two sources and is therefore less constraining for maser models, the fast rise time and overall shape of its light curve are consistent with SR and warrant a close examination within its context.
\begin{center}
   \begin{figure}
        \includegraphics[width=\columnwidth]{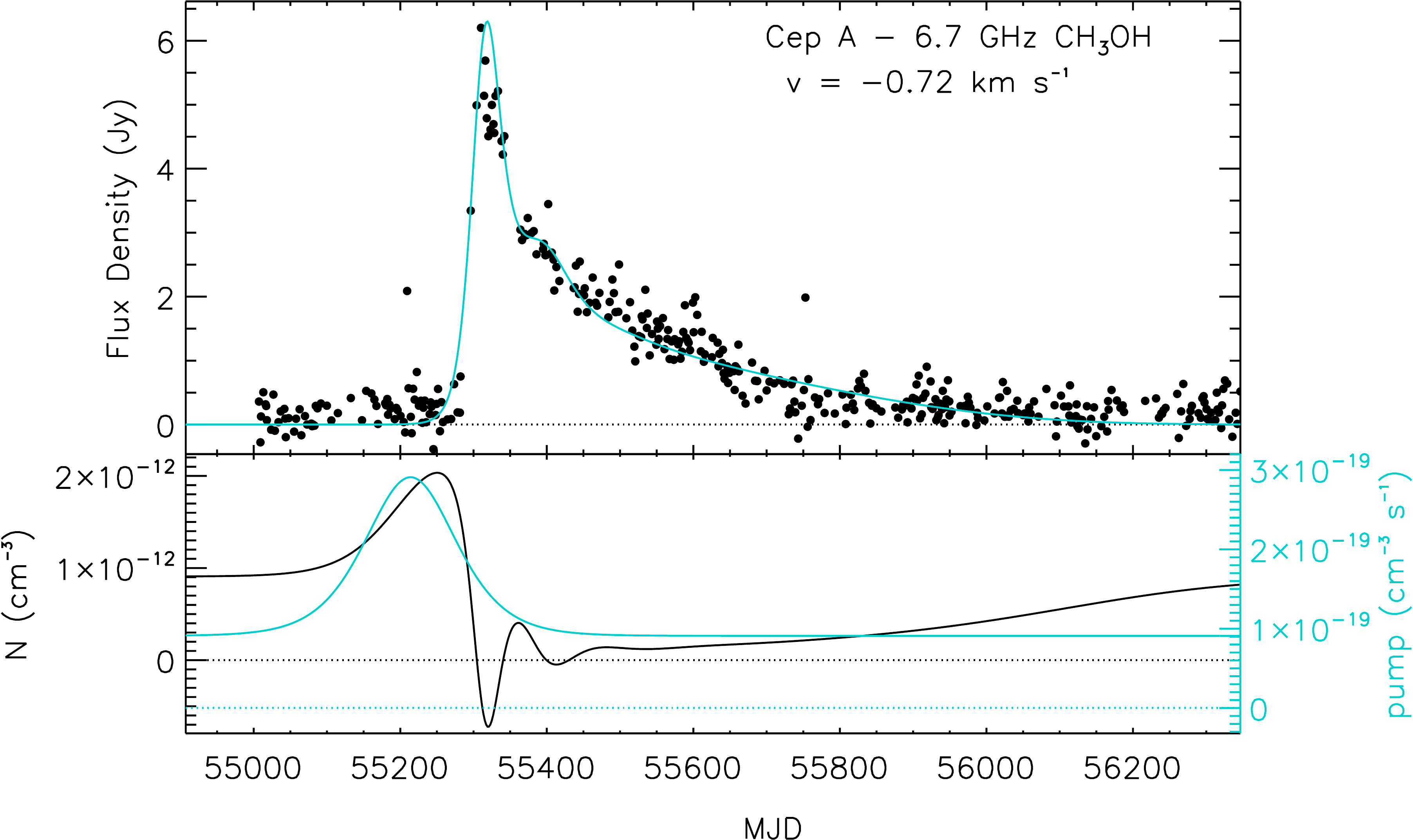}
        \caption{Same as Figure \ref{fig:S255FitPump} but for Cep A at $v_{\mathrm{lsr}} = -0.72~\mathrm{km~s}^{-1}$. The SR fit is produced using a single SR sample of length $L = 125$~au composed of $N_\mathrm{SR} \sim 3 \times 10^{19}$ inverted methanol molecules. The time-scales for relaxation and dephasing are $T_1 = 9.99 \times 10^6$~s and $T_2 = 2.72 \times 10^6$~s, respectively. Similar to S255-NIRS3, the inversion level prior to the arrival of the pump pulse corresponds to $0.1~\mathrm{cm}^{-3}$ for a molecular population spanning a velocity range of $1~\mathrm{km~s}^{-1}$ is assumed. The SR fit yields an inverted column density $\left(nL\right)_{\mathrm{SR}} = 3.8 \times 10^{3}$~cm$^{-2}$. The flux density of the SR model is scaled to the data.}
        \label{fig:CepAFitPump}
    \end{figure}
\end{center}

In Figure \ref{fig:CepAFitPump}, we focus on the $v_{\mathrm{lsr}} = -0.72~\mathrm{km~s}^{-1}$ channel to detail the SR model's fit parameters. As in Figures \ref{fig:S255FitPump} and \ref{fig:G24p329FitPump}, the evolution of the inverted population density (black curve, bottom panel; vertical axis on the left), the total pump rate (cyan curve, bottom panel; vertical axis on the right) and the SR model fit (cyan curve, top panel) superposed on the data (dots) are given as a function of retarded time. The SR model fit is produced using an SR sample of length $L =125$~au, where $N_\mathrm{SR} = 3.2 \times 10^{19}$ excited methanol molecules cooperatively produce an integrated flux density of $5.94\times 10^{-34}~\mathrm{erg~s}^{-1}\mathrm{cm}^{-2}$ (or a flux density of $5.94 \times 10^{-4}$ Jy for a bandwdith of $\sim10^{-7}$~Hz) at a distance of 0.7~kpc. As for S255IR-NIRS3 and G24.329+0.144, we started with an initial inverted population density of $0.1~\mathrm{cm}^{-3}$ for a molecular population spanning a velocity range of $1~\mathrm{km~s}^{-1}$, which through the fitting process resulted in an inversion level $\hat{\mathbb{N}}_0 = 9.09 \times 10^{-13}$~cm$^{-3}$ prior to the arrival of the pump pulse (i.e., for a single SR sample). Accordingly, the constant pump rate was set to $7.15 \times 10^{-19}$~cm$^{-3}$s$^{-1}$. The relaxation and dephasing time-scales were set to $T_1 = 9.99\times 10^6$~s and $T_2 = 2.72\times 10^6$~s, respectively. 

After the arrival of the pump pulse of amplitude $\Lambda_1 = 2 \times 10^{-19}$ cm$^{-3}$s$^{-1}$ and duration $T_{\mathrm{p}} = 7.21 \times 10^6$~s, the inversion level grows until it reaches approximately twice its initial value, at which point SR is initiated. We once again note the fast rise time of the burst in comparison to the duration of the pump pulse. Our SR model yields an inverted population column density $\left(nL\right)_\mathrm{SR} = 3.8 \times 10^{3} $ cm$^{-2}$. As before, our model can accommodate changes in $n$ and $L$ by keeping $\left(nL\right)_\mathrm{SR}$ unaltered. The radius of the cylindrical SR sample was found to be $5.17\times 10^7$~cm (or $3.45\times 10^{-6}$~au).

The SR model fits to the other velocity channels in Figure \ref{fig:CepAFit} are produced using the same inverted population density, relaxation time-scale $T_1$, and dephasing time-scale $T_2$. Similarly, pump rate parameters were kept the same. The only variations in the fitting process were slight changes in the SR sample length $L$, which ranged between 125~au to 129~au. 

%%\textbf{Finally, we note that our SR models for S255IR-NIRS3, Cepheus A, and G24.329+0.144 yield inverted column densities $\sim 10^3-10^4~\mathrm{cm}^{-2}$ consistent with results obtained in previous SR analyses. More precisely, our SR models of the star-forming regions G107.298+5.639 and G33.64-0.21 for the methanol 6.7 GHz line revealed similar inverted column densities $\sim 10^4~\mathrm{cm}^{-2}$ and time-scales \citep{Rajabi2017}.}

%%%%%%%%#################   Conclusion  ########################
%%%%%%%%########################################################

\section{Conclusions}\label{sec:conclusion}
We applied Dicke's SR model to the methanol 6.7 GHz spectral line often associated with masers in star-forming regions. The build-up of coherent interactions in SR requires population inversion, velocity coherence among emitters, and long dephasing/relaxation time-scales relative to the time-scale of SR. The large number of 6.7 GHz methanol masers detected in star-forming regions emphasizes that the first two requirements can readily be satisfied in the ISM. This, and the previous evidence found in G107.298+5.639 and G33.64-0.21 \citep{Rajabi2017}, motivated a search for SR in masing regions. In this paper, we presented new evidence for SR in three different star-forming regions: S255IR-NIRS3, G24.329+0.144, and Cepheus A. Our analysis showed that some of the flux-density flares exhibiting fast flux rise times and asymmetric light curves reported in these sources can naturally be explained within the context of SR. We discussed how a transition from a maser mode with an intensity $I \propto N$, where $N$ is the number of excited molecules, to an SR mode with an intensity $I_{\mathrm{SR}} \propto N^2$ is key to understanding fast flux rise times and asymmetric light curves. SR, as a more efficient energy release mechanism, manifests itself through strong bursts of radiation emanating from spatially compact regions. This transition from maser to SR radiation modes can take place when a threshold set by a critical inverted column density is exceeded. This can happen through a change in radiative pumping, which can lead to elevated inverted population densities and the initiation of SR. We find that SR samples containing $10^{19}-10^{20}$ excited and entangled methanol molecules with inverted population column densities ranging from  $10^3$ to $10^4~\mathrm{cm}^{-2}$ can reproduce the detected flare profiles.

The study of superradiance in astrophysics is admittedly in its beginnings. More work is needed, both theoretically and observationally, to get a better picture and characterisation of the different environments that can support this effect. In particular, extreme resolution VLBI observations will be an excellent way of testing our SR flare mechanism and model.

%Column densities:G107-methanol source fit: TR=2.1 hours, Tprime= 90 TR and nL= 3.5 times 10 ^ 4 cm^{-2} and water nL = 8.4 times 10 ^ 4 cm^{-2}, TR water= 7.7 hrs, Tprime water= 

%%%Methanol 6.7 GHz Flares in G33.64-0.21.----nL = 7 times 10 ^ 4 cm^{-2}, TR= 1.1 hour, Tprime = 600 TR

%% Water 22 GHz Mattila --Water 22 GHz Flares in Cepheus A-- TR = 8.2 hours, Tprime = 700 TR--nL = 6 times 10 ^ 4 cm^{-2}

%%% The U Orionis OH 1612 MHz, Jewell data---TR = 6.5 days, Tpirme = 393 TR- nL = 3.4 times 10 ^ 3 cm^{-2}?--L =3.4 times 10 ^ 4 cm 

%% IRAS18276-1431 (OH17.7-2.0) - OH 1612 MHz- Wolak data--TR = 42 days, Tprime = 61 TR--nL=5.2 times 10^2 cm^{-2}?, L= 5.2 times 10^3 cm
%%%%%%%%############################################################
%%%%%%%%############################################################
\section*{Acknowledgements}
M.H.'s research is funded through the Natural Sciences and Engineering Research Council of Canada Discovery Grant RGPIN-2016-04460.
A.B., M.O., M.S., and P.W. acknowledge partial support from the National Science Centre of Poland under grant 2016/21/B/ST9/01455.

%%%%%%%%%%%%%%%%%%%%%%%%%%%%%%%%%%%%%%%%%%%%%%%%%%
%%%%%%%%%%%%%%%%%%%% REFERENCES %%%%%%%%%%%%%%%%%%
% The best way to enter references is to use BibTeX:
\bibliographystyle{mnras}
\bibliography{SR2-bib} 

%%%%%%%%%%%%%%%%%%%%%%%%%%%%%%%%%%%%%%%%%%%%%%%%%%
%%%%%%%%%%%%%%%%% APPENDICES %%%%%%%%%%%%%%%%%%%%%
% \appendix
% \section{}
%%%%%%%%%%%%%%%%%%%%%%%%%%%%%%%%%%%%%%%%%%%%%%%%%%

% Don't change these lines
\bsp	% typesetting comment
\label{lastpage}
\end{document}